\documentclass[12pt]{article}
\usepackage{amsmath,latexsym}
\usepackage{graphicx}
\begin{document}

 \title{\Huge Thin Shell Wormhole in Heterotic String Theory}
 \author{F.Rahaman$^*$, M.Kalam$^{\ddag}$ and  S.Chakraborti$^*$  }
\date{}
 \maketitle

 \begin{abstract}
Using ' Cut and Paste ' technique, we develop a thin shell
wormhole in heterotic string theory. We determine the surface
stresses, which are localized in the shell, by using
Darmois-Israel formalism. The linearized stability of this thin
wormhole is also analyzed.
\end{abstract}

  \footnotetext{ Pacs Nos :  04.20 Gz,04.50 + h, 04.20 Jb   
\\
 $*$Dept.of Mathematics, Jadavpur University, Kolkata-700 032,
 India:
                                  E-Mail:farook\_rahaman@yahoo.com\\
$\ddag$Dept. of Phys. , Netaji Nagar College for Women, Regent Estate, Kolkata-700092, India.\\

}
    \mbox{} \hspace{.2in}

\title{\Huge Introduction: }

Recently, several Physicists have given their attention to develop
thin shell wormhole Models. Visser, the pioneer, presented this
new class of wormholes by using 'Cut and Paste' technique [1]. The
model is constructed by surgically grafting two Schwarzschild
spacetimes together in such a way that no event horizon is
permitted to form . Taking  two copies of region from
Schwarzschild geometry with $ r\geq a$ : $ M^\pm \equiv ( x | r
\geq a ) $ where $ a > r_h $ [ $r_h$ = radius of event horizon and
$ \pm $ indicates two copies ], one can paste two copies $ M^\pm $
together at the hypersurface $ \Sigma=\Sigma^\pm = ( x | r = a ) $
results in a geodesically complete regions to make a thin shell
wormhole. This construction creates a geodesically complete
manifold $ M = M^+ \bigcup M^- $ with two asymptotically flat
regions connected by a throat placed at $ \Sigma$. In this case,
the surface energy density of the shell is found to be negative
(one can use Israel-Darmois [2] formalism to prove it). This
indicates  the presence of exotic matter in the throat. After some
years of this proposal, Visser with his collaborator Poisson had
analyzed the stability of thin shell wormhole constructed by
joining the two Schwarzschild geometries [3]. Ishak and Lake have
examined the stability of transparent spherical symmetric
thin-shell wormholes [4] .

Eiroa and Romero [5] have studied the linear stability of charged
thin shell wormholes constructed by joining the two
Reissner-Nordstr\"{o}m spacetimes under spherically symmetric
perturbations.  Lobo and Crawford [6] have extended the linear
stability analysis to the thin shell
wormholes with Cosmological Constant. \\
Eiroa and Simeone [7] have studied cylindrically symmetric thin
shell wormhole geometries associated to gauge cosmic strings.
Also, the same authors have constructed a charged thin shell
wormhole in dilaton gravity and they have shown that the
reduction of the total amount of exotic matter is dependent on
the Dilaton-Maxwell coupling parameter [8]. The five  dimensional
thin shell wormholes in Einstein-Maxwell theory with a Gauss
Bonnet term has been studied by Thibeault et al [9]. They have
made a linearized stability analysis under radial perturbations.
Recently, the present authors have studied thin shell wormholes in
higher dimensional Einstein-Maxwell theory which is constructed by
Cutting and Pasting two metrics corresponding to a higher
dimensional Reissner-Nordstr\"{o}m black hole [10].\\
Super string theory plays an important role to unify gravity with
all other fundamental interactions in nature. In recent past, the
general dyonic black hole solutions to the heterotic string theory
were obtained by Jhatkar et al [11] and Cvetic et al [12]. They
had given a general black hole solution to the Heterotic string
compactified on a four dimensional torus. These black hole
solutions exhibit several different properties compared to the
Reissner-Nordstr\"{o}m black holes. After the publication of the
paper by Visser [1], there has been a growing interest in the
study of thin shell wormholes constructed by black holes. And
since, the charged black holes of general relativity and string
theory are qualitatively different, it is expected that thin shell
wormhole constructed by black holes in string theory will give new
features. The purpose of  this article is to study thin shell
wormhole in heterotic string theory. We develop the model by
cutting and pasting two metrics corresponding to a four
dimensional dyonic black hole solution of toroidally compactified
heterotic string theory. The linearized stability is analyzed
under radial perturbations around static solutions. \\ The paper
is organized as follows: The black hole solution in heterotic
string theory is rewritten in section 2. In section 3, thin shell
wormhole has been constructed by means of the Cut and Paste
tactics. A linearized stability analysis is studied in section 4.
In section 5, the total amount of exotic matter has been
calculated. Section 6 is devoted to a brief summary and
discussion.

\pagebreak

\title{\Huge2.Dyonic black holes in
heterotic string theory }

Following [11-13], we consider the metric which can be associated
to a four dimensional Dyonic black hole solution of toroidally
compactified heterotic string theory as

\begin{equation}
               ds^2=  - f(r) dt^2+ \frac{dr^2}{ f(r)} + h(r) d\Omega_2^2
               \end{equation}
where
\begin{equation}
               f(r) = \frac{(r+b)(r-b)}{h(r)}
               \end{equation}
\begin{equation}
               h^2(r) = (r+\hat{Q_1})(r+\hat{Q_2})(r+\hat{P_1})(r+\hat{P_2})
               \end{equation}
\begin{equation}
               \hat{Q_1} = \sqrt{Q_1^2 + b^2}
               \end{equation}

           etc..\\

The above black hole consists with two independent electric
charges $Q_1$ and $Q_2$ ( same signs ) as well as two independent
magnetic charges $P_1$ and $P_2$. The position of the horizon is
specified by the real positive parameter 'b'. \\
For the sake of simplicity, we do not include the 28 gauge
fields to which the 28 components of charges are coupled.\\
The dilaton field takes the form [13]
\begin{equation}
               e^{2\phi} =
               \frac{(r+\hat{P_1})(r+\hat{P_2})}{(r+\hat{Q_1})(r+\hat{Q_2})}
               \end{equation}
The above expression indicates that the dilaton field does not in
general vanish for these solutions.  It tends to zero
asymptotically as $ r \rightarrow \infty $ .

When $ Q_i = P_i = 0 $, then the metric (1) reduces to
Schwarzschild  geometry:\\

         $ ds^2=  - (1-\frac{2b}{R}) dt^2+ (1-\frac{2b}{R})^{-1}dR^2 + R^2
d\Omega_2^2 $ \\

While for $ Q =  Q_i = P_i \neq 0 $, one can obtain \\

$ ds^2=  - (1-\frac{2\sqrt{Q^2+b^2}}{R}+\frac{b^2}{R^2}) dt^2+
(1-\frac{2\sqrt{Q^2+b^2}}{R}+\frac{b^2}{R^2})^{-1}dR^2 + R^2
d\Omega_2^2 $, \\

which corresponds to Reissner-Nordstr\"{o}m black hole geometry.

\title{\Huge3. The Darmois-Israel formalism
                and Cut and Paste construction: }

As we are interested to construct thin shell wormhole in
heterotic string theory by using Cut and Paste technique, from
the geometry (1), we take two copies of the region with $ r\geq
a$: $ M^\pm = ( x \mid r \geq a )  $ where $ a \geq r_h = b $ (
position of event horizon ) and paste the two pieces together  at
the hypersurface $ \Sigma = \Sigma^\pm = ( x \mid r = a )$.

This new construction results in a geodesically complete manifold
$ M = M^+ \bigcup M^- $ with a matter shell at the surface $ r =
a $ , where the throat of the wormhole is located. Thus a single
manifold M is obtained which connects two asymptotically flat
regions at their boundaries $\Sigma$ and the throat is placed at
$\Sigma$(here $\Sigma$ is a synchronous time like hypersurface).
Following Darmois-Israel formalism, we shall determine the
surface stresses at the junction boundary. The intrinsic
coordinates in $\Sigma$ are taken as $ \xi^i = ( \tau, {\theta},
\phi)$  with $\tau$ is the proper time on the shell. To understand
the dynamics of the wormhole, we assume  the radius of the throat
be a function of the proper time $ a = a(\tau)$. The parametric
equation for $\Sigma$ is defined   by

\begin{equation}\Sigma : F(r,\tau ) = r - a(\tau)\end{equation}

The second fundamental form ( extrinsic curvature ) associated
with the two sides of the shell are
\begin{equation}K_{ij}^\pm =  - n_\nu^\pm\ [ \frac{\partial^2X_\nu}
{\partial \xi^i\partial \xi^j } +
 \Gamma_{\alpha\beta}^\nu \frac{\partial X^\alpha}{\partial \xi^i}
 \frac{\partial X^\beta}{\partial \xi^j }] |_\Sigma \end{equation}
where $ n_\nu^\pm\ $ are the unit normals to $\Sigma$ in M:
\begin{equation} n_\nu^\pm =  \pm   | g^{\alpha\beta}\frac{\partial F}{\partial X^\alpha}
 \frac{\partial F}{\partial X^\beta} |^{-\frac{1}{2}} \frac{\partial F}{\partial X^\nu} \end{equation}
[ $i,j = 1,2,3$ corresponding to boundary $\Sigma$ ; $ \alpha,
\beta = 1,2,3,4$ corresponding to original spacetime ]
 with $ n^\mu n_\mu = 1 $.

 The intrinsic metric at $\Sigma$ is given by
\begin{equation}
               ds^2 =  - d\tau^2 + a(\tau)^2 d\Omega^2
               \end{equation}
The position of the throat of the wormhole is described by $ X^\mu
= ( t, a(t), \theta, \phi ) $. The unit normal to $\Sigma$ is
given by
\begin{equation}
               n^\nu  =  ( \frac{\dot{a}}{f(a)} , \sqrt{(f(a) +
               \dot{a}^2)}, 0, 0 )
               \end{equation}

               \pagebreak

Now using equations (1), (7) and (10), the non trivial components
of the extrinsic curvature are given by
\begin{equation}
                K_{\tau\tau}^{\pm}  =  \mp \frac{\frac{1}{2}f^{\prime}(a) +
                \ddot{a}}
                {\sqrt{(f(a) +
               \dot{a}^2)}}
              \end{equation}
              \begin{equation}
                K_{\theta\theta}^{\pm}= K_{\phi\phi}^{\pm}
                  =  \pm \frac{h^{\prime}(a)}{2h(a)} \sqrt{(f(a) +
               \dot{a}^2)}
              \end{equation}
 We define jump of the discontinuity of the extrinsic
curvature of the two sides of $\Sigma $ as $[ K_{ij} ] = K_{ij}^+
- K_{ij}^-$ and $ K = [K_i^i ] = trace [K_{ij}]$.

The Ricci tensor at the throat can be calculated in terms of the
discontinuity of the second fundamental forms ( extrinsic
curvature ). This jump discontinuity, together with Einstein field
equations, provides the stress energy tensor of $\Sigma $, where
throat is localized: $T^{\mu\nu}  = S^{\mu\nu}\delta(\eta) $ [
$\eta$ denotes the proper distance away from the throat ( in the
normal direction )] with,
\begin{equation}S_j^i  =  - \frac{1}{8\pi} ( [K_j^i]  - \delta_j^i K )\end{equation}
where
 $ S_j^i = diag ( - \sigma , p_{\theta}, p_{\phi}) $
 is the  surface  energy tensor with $\sigma$ , the surface density and $p_\theta$ and $p_\phi$,
  the
surface pressures.

Now taking into account the equation (13), one can find

\begin{equation}
               \sigma =  - \frac{1}{4\pi }\frac{h^\prime(a)}{h(a)}\sqrt{f(a) + \dot{a}^2}
               \end{equation}

\begin{equation}
              p_{\theta} = p_{\phi} =
  p =  \frac{1}{8\pi }\frac{h^\prime(a)}{h(a)}\sqrt{f(a) + \dot{a}^2} + \frac{1}{8\pi
}\frac{2\ddot{a} + f^\prime(a) }{\sqrt{f(a) + \dot{a}^2}}
               \end{equation}

[ over dot and prime mean, respectively, the derivatives with
respect to $\tau$ and a ]

 From equations (14) and (15), one can
verify the energy conservation equation:

\begin{equation}
               \frac {d}{d \tau} (\sigma h(a)) + p \frac{d}{d \tau}(h(a))= 0
               \end{equation}
or
\begin{equation}
               \dot{\sigma} +  \frac{\dot{h}(a)}{h(a)}( p + \sigma ) = 0
               \end{equation}

The first term represents the variation of the internal energy of
the throat and the second term is the work done by the throat's
 internal forces. Negative energy density in (14) implies  the
existence of exotic matter at the shell.

\title{\Huge4. Linearized Stability Analysis: }

Rearranging equation (14), we obtain the thin shell's  equation of
motion

            \begin{equation}  \dot{a}^2 + V(a)= 0  \end{equation}

                Here  the potential is defined  as
\begin{equation}
              V(a) =  f(a) - [\frac{4\pi
              h(a)\sigma(a)}{h^{\prime}(a)}]^2
                 \end{equation}

\title{\Huge4.1  Static Solution: }

The above single dynamical equation (18) completely determines the
motion of the wormhole throat. One can consider a linear
perturbation around a static solution with radius $a_0$. We are
trying to find a condition, for which stress energy tensor
components at $a_0$ will obey null energy condition. For a static
configuration of radius $a_0$, we obtain respective values of the
surface energy density and the surface pressure by using the
explicit form (1) of the metric as
\begin{equation}
              \sigma_0 =  -
              \frac{1}{8\pi}\frac{\sqrt{(a_0+b)(a_0-b)}}{[(a_0+\hat{Q_1})
              (a_0+\hat{Q_2})(a_0+\hat{P_1})(a_0+\hat{P_2})]^{\frac{1}{4}}}[
\frac{1}{(a_0+\hat{Q_1})}+\frac{1}{
              (a_0+\hat{Q_2})}+\frac{1}{(a_0+\hat{P_1})}+\frac{1}{(a_0+\hat{P_2})}]
                 \end{equation}
\begin{equation}
              p_0 =
              \frac{1}{8\pi}\frac{\sqrt{(a_0+b)(a_0-b)}}{[(a_0+\hat{Q_1})
              (a_0+\hat{Q_2})(a_0+\hat{P_1})(a_0+\hat{P_2})]^{\frac{1}{4}}}[
\frac{2a_0}{a_0^2 - b^2}]
                 \end{equation}

One can see that surface energy density is always negative,
implying the violation of weak and dominating energy conditions.
The null energy condition will satisfy if $ \sigma_0 +  p_0 > 0 $
i.e.

$  \frac{1}{8\pi}\frac{\sqrt{(a_0+b)(a_0-b)}}{[(a_0+\hat{Q_1})
              (a_0+\hat{Q_2})(a_0+\hat{P_1})(a_0+\hat{P_2})]^{\frac{1}{4}}}[
\frac{2a_0}{a_0^2 - b^2}- ( \frac{1}{(a_0+\hat{Q_1})}+\frac{1}{
              (a_0+\hat{Q_2})}+\frac{1}{(a_0+\hat{P_1})}+\frac{1}{(a_0+\hat{P_2})})]> 0 $

Thus null energy condition is obeyed if
\begin{equation}  \frac{2a_0}{a_0^2 - b^2} >  ( \frac{1}{(a_0+\hat{Q_1})}+\frac{1}{
              (a_0+\hat{Q_2})}+\frac{1}{(a_0+\hat{P_1})}+\frac{1}{(a_0+\hat{P_2})})
                 \end{equation}
In particular if $ Q_i = P_i = 0 $, then inequality (22) implies $
a_0 < 2b = r_h $( the position of the event horizon of
Schwarzschild black hole). Hence for the thin shell wormhole
constructed by joining the Schwarzschild geometries, the null
energy condition is always violated ( as we have considered $ a_0
> 2b = r_h $ ).

Also, if we take $ Q = Q_i = P_i \neq 0 $, then inequality (22)
implies

$ a_0 < \frac{\sqrt{Q^2 + b^2}}{2} + \frac{\sqrt{Q^2 + 9b^2}}{2}$.

When, $ Q = Q_i = P_i \neq 0 $, the metric (1) reduces to the
Reissner-Nordstr\"{o}m geometry having the position of event
horizon $ r_h = \sqrt{Q^2 + b^2} + 2Q^2 $.

Since we have considered $ a_0 > r_h $, then the null energy
energy condition is satisfied if

$ \sqrt{Q^2 + b^2} + 2Q^2  < a_0 < \frac{\sqrt{Q^2 + b^2}}{2} +
\frac{\sqrt{Q^2 + 9b^2}}{2}$.

\title{\Huge4.2  Stability Analysis: }

 Linearizing around a static solution situated at $a_0$,
one can expand V(a) around $a_0$ to yield
\begin{equation}
              V =  V(a_0) + V^\prime(a_0) ( a - a_0) + \frac{1}{2} V^{\prime\prime}(a_0)
              ( a - a_0)^2 + 0[( a - a_0)^3]
                 \end{equation}
where prime denotes derivative with respect to $a$.

Since we are linearizing around a static solution at $ a = a_0 $,
we have $ V(a_0) = 0 $ and $ V^\prime(a_0)= 0 $. The stable
equilibrium configurations correspond to the condition $
V^{\prime\prime}(a_0)> 0 $. Now we define a parameter $\beta$,
which is interpreted as the speed of sound, by the relation
\begin{equation}
              \beta^2(\sigma) = \frac{ \partial p}{\partial
              \sigma}|_\sigma
                 \end{equation}
Using conservation equation (16), we have

$ V^{\prime\prime}(a) = f^{\prime\prime} - \frac{32\pi^2
h^2(\sigma^\prime)^2}{(h^\prime)^2 }
  - \frac{128\pi^2 a \sigma \sigma^\prime h}{h^\prime}
   - \frac{32\pi^2 h^2\sigma}{(h^\prime)^2} [ (\frac{(h^\prime)^2}{h^2} -
   \frac{h^{\prime\prime}}{h})( p + \sigma ) -
 \frac{h^\prime}{h} \sigma^{\prime} ( 1 + \beta^2) ]
  + \frac{128\pi^2 \sigma \sigma^\prime h^2h^{\prime\prime}}{(h^\prime)^3}
  + \frac{96\pi^2 \sigma^2 hh^{\prime\prime}}{(h^\prime)^2}
  + \frac{32\pi^2 \sigma ^2 h^2h^{\prime\prime\prime}}{(h^\prime)^3}
  - \frac{96\pi^2\sigma^2 h^2(h^{\prime\prime})^2}{(h^\prime)^4}
  - 32\pi^2\sigma^2 $

The wormhole solution is stable if $ V^{\prime\prime}(a_0)> 0 $
i.e. if
\begin{equation}
              \beta_0^2 <
              [\frac{\frac{5h_0^{\prime\prime}f_0}{h_0} +
               \frac{h_0^{\prime}f_0^{\prime}}{h_0}+  \frac{(f_0^{\prime})^2}{2f_0}
               - \frac{2h_0^{\prime\prime\prime}f_0}{h_0^\prime} -
               \frac{h_0^{\prime\prime}f_0^\prime}{h_0^\prime} - \frac{3(h_0^{\prime})^2f_0}
               {h_0^2} - f_0^{\prime\prime}}{\frac{2h_0^{\prime\prime}h_0^\prime \sqrt{f_0}}
               {h_0^2} + \frac{(h_0^{\prime})^2f_0^\prime}{h_0^2 \sqrt{f_0}}
          - \frac{(h_0^{\prime})^3\sqrt{f_0}}
               {h_0^3}} ] -1
                 \end{equation}
or
\begin{equation}
              \beta_0^2 < \frac{ A + B + C - D - E  - G - H}{M + N - L} - 1
                 \end{equation}
                 where A, B, C, D, E, G, H, M, N, L are given
                 in the appendix.

Thus if one treats $a_0$, b, $Q_i$ and $P_i$   are specified
quantities, then the stability of the configuration requires the
above restriction on the parameter $\beta_0$. This means there
exists some part of the parameter space, where the throat location
is stable. The dependence of the regions of stability will vary
with different choices of the parameters.

\pagebreak

\title{\Huge5. Total amount of exotic matter: }

To characterize the viability of traversable wormhole, it is
important to quantify the total amount of exotic matter. Now, we
shall determine the total amount of exotic matter for the thin
wormhole in heterotic string theory. The total amount of exotic
matter can  be quantified by the integrals ( In this case radial
pressure $ p_r = 0 $ and we have $ \sigma <0 , \sigma + p_r <0$
i.e. both energy conditions are violated. The transverse pressure
is $ p = p_t $ and one can see from (22) that the sign of $ \sigma
+ p_t $ is not fixed but depends on the value of the parameters )

\begin{equation}
             \Omega =  \int [\sigma + p_r]
\sqrt{-g}d^3x
                 \end{equation}

Following Eiroa and Simone [8-9] , we introduce a new radial
coordinate $ R  =  \pm ( r -a ) $ in M ( $\pm $ for $M^{\pm}$
respectively ) so that

\begin{equation}
            \Omega =  \int_0^{2\pi} \int_0^{\pi}\int_{-\infty}^\infty [\sigma  + p_r]
\sqrt{-g}dRd{\theta}d{\phi}
                 \end{equation}

Since the shell does not exert radial pressure and the energy
density is located on a thin shell surface, so that $ \sigma  =
\delta(R)\sigma_0$, then we have

$             \Omega = \int_0^{2\pi} \int_0^\pi  [\sigma \sqrt{-g}
]|_{r=a_0} d{\theta}d{\phi} = 4\pi h(a_0)\sigma(a_0)$.

Thus one gets,
$
             \Omega =  - \frac{1}{2} \sqrt{( a_0 + b)(a_0 - b)}
[(a_0+\hat{Q_1})
              (a_0+\hat{Q_2})(a_0+\hat{P_1})(a_0+\hat{P_2})]^{\frac{1}{4}}
                            [ \frac{1}{(a_0+\hat{Q_1})}+\frac{1}{
              (a_0+\hat{Q_2})}+\frac{1}{(a_0+\hat{P_1})}+\frac{1}{(a_0+\hat{P_2})}]
             $

Now, we shall consider two cases (i) the variation of $ \Omega $
with respect to b (ii) the variation of $ \Omega $ with respect to
Q , where Q is understood to include both electric and both
magnetic charges.

Case-I :

For a lot of useful information, the variation of
$\frac{\Omega}{a_0}$ with $\frac{b}{a_0}$ for different values of
charges ( electrical and magnetic ) is depicted in the fig-I.

\begin{figure}[htbp]
    \centering
        \includegraphics[scale=.8]{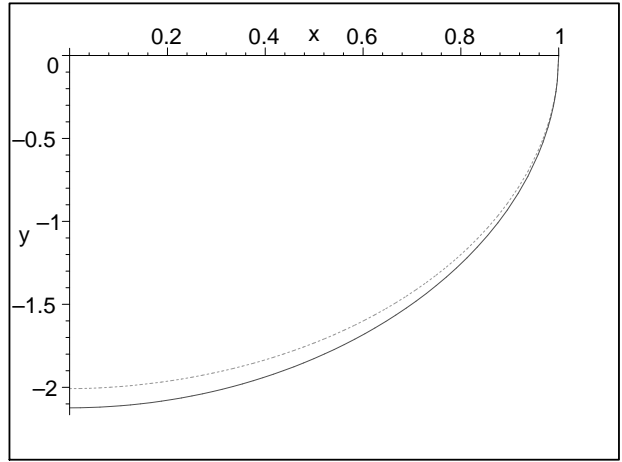}
        \caption{ We choose two cases as : $ \frac{Q^2_1}{a^2_0}  = 1$ , $ \frac{Q^2_2}{a^2_0}  = 2^2$,
        $ \frac{P^2_1}{a^2_0}  = 3^2$, $ \frac{P^2_2}{a^2_0}  =
        4^2$ (for solid line) and $ \frac{Q^2_1}{a^2_0}  = (.1)^2$ , $ \frac{Q^2_2}{a^2_0}  = (.2)^2$,
        $ \frac{P^2_1}{a^2_0}  = (.3)^2$, $ \frac{P^2_2}{a^2_0}  =
        (.4)^2$( for dotted line ).
        We define
        $  \frac{b}{a_0} = x $ ,  $\frac{\Omega}{a_0} = y $ and plot y Vs.  x .
        The variation of total amount of exotic matter on the shell
        with respect to the parameter b is shown in the figure.}
    \label{fig:I}
\end{figure}

\pagebreak
If $ a_0, P_i,  Q_i >> b $, then

$ \Omega =  - \frac{1}{2}  a_0  [(a_0+{Q_1})
              (a_0+{Q_2})(a_0+{P_1})(a_0+{P_2})]^{\frac{1}{4}}
                            [ \frac{1}{(a_0+{Q_1})}+\frac{1}{
              (a_0+{Q_2})}+\frac{1}{(a_0+{P_1})}+\frac{1}{(a_0+{P_2})}]$

Also, if one considers, $ P_i = Q_i = Q $, then
 $\Omega =  - 2 \sqrt{( a_0 + b)(a_0 - b)}$.

 The variation of $\frac{\Omega}{a_0}$ with respect to the parameter $\frac{b}{a_0}$ is
shown in  fig-II.
\begin{figure}[htbp]
    \centering
        \includegraphics[scale=.8]{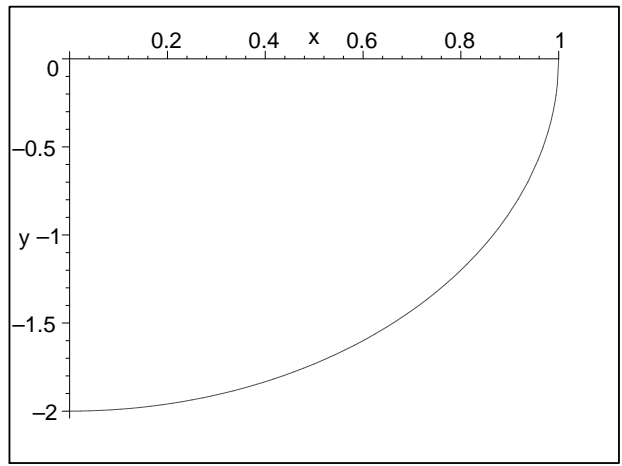}
        \caption{ We define
        $  \frac{b}{a_0} = x $ ,  $\frac{\Omega}{a_0} = y $ and plot y Vs.  x .
        For the case, $ P_i = Q_i = Q $,  the variation of total amount of exotic matter on the shell
        with respect to the parameter b is shown in the figure.}
    \label{fig:II}
\end{figure}

The above equations describing $\Omega$ indicate that the total
amount of exotic matter can be reduced by taking the wormhole
radius close to parameter 'b'.

Case-II :

In this case, we shall treat b as a fixed constant. Now,for large
Q, the expression of $\Omega$ will take the following form,

$ \Omega =  - \frac{1}{2} \sqrt{( a_0 + b)(a_0 - b)} [ \prod
{Q^{\frac{1}{4}}}( 1 + \frac{a_0}{4Q} + \frac{4b^2 - 3a_0^2}{Q^2}
+ 0(\frac{1}{Q^3})) ] [ \sum ( \frac{1}{Q} - \frac{a_0}{Q^2} +
0(\frac{1}{Q^3})) ]
 $

The above expression indicates that the total amount of exotic
matter will be needed more with the increase of Q. Thus required
exotic matter is minimum, when Q takes its minimum values, say $ Q
= Q_{min}$ .

\title{\Huge6. Concluding remarks: }

In recent years, string theory has become an  active area of
research because it generalizes the Einstein theory in many ways.
It is of interest to investigate how the properties of black holes
are modified when the black holes of the effective four
dimensional heterotic string theory compactified on a six torus
are considered [14-15]. So black hole solution arising from
toroidally compactified heterotic string theory has been an
intriguing subject for researchers. In this article we have
constructed a new class of thin wormhole by surgically grafting
two black hole spacetimes arising from heterotic string theory on
a six torus. We have given our attention  only  how to get the
geometry of the thin wormhole. We have analyzed the dynamical
stability of the thin shell, considering linearized radial
perturbations around stable solution. To analyze this, we define a
parameter $ \beta^2 = \frac{p^\prime }{\sigma^\prime} $ as a
parametrization of the stability of equilibrium. We have obtained
a restriction on $\beta^2$ to get stable equilibrium of the thin
wormhole( see eq.(26)). It is shown that the matter within the
shell violates the weak energy condition but that matter may obey
null energy condition. Although it is not possible to provide the
mechanism that provide the exotic matter but rather we have
calculated an integral  measuring of the total amount of exotic
matter. Finally,  we have shown that total amount of exotic matter
needed to support traversable wormhole can be reduced as desired
with the suitable choice of the parameters.

\pagebreak

 {  \Huge Acknowledgments }

          F.R. is thankful to Jadavpur University and DST , Government of India for providing
          financial support under Potential Excellence and Young
          Scientist scheme . MK has been partially supported by
          UGC,
          Government of India under Minor Research Project scheme.
          We are also grateful to the anonymous referee for his
valuable comments and constructive suggestions.\\

\pagebreak

{ \bf \Huge Appendix }
 \centering

 $ A =   5 [(\frac{-1}{2(a_0+\hat{Q_1})^2}+\frac{-1}{
              2(a_0+\hat{Q_2})^2}+\frac{-1}{2(a_0+\hat{P_1})^2}+\frac{-1}{2(a_0+\hat{P_2})^2})
              + \frac{1}{4}(\frac{1}{(a_0+\hat{Q_1})}+\frac{1}{
              (a_0+\hat{Q_2})}+\frac{1}{(a_0+\hat{P_1})}+\frac{1}{(a_0+\hat{P_2})})^2]
              [\frac{(a_0+b)(a_0-b)}{[(a_0+\hat{Q_1})(a_0+\hat{Q_2})(a_0+\hat{P_1})
              (a_0+\hat{P_2})]^{\frac{1}{2}}}] $

$ B = [\frac{1}{2(a_0+\hat{Q_1})}+\frac{1}{
              2(a_0+\hat{Q_2})}+\frac{1}{2(a_0+\hat{P_1})}+\frac{1}{2(a_0+\hat{P_2})}
              ][\frac{2a_0}{a_0^2-b^2}-\frac{1}{2}(\frac{1}{(a_0+\hat{Q_1})}+\frac{1}{
              (a_0+\hat{Q_2})}+\frac{1}{(a_0+\hat{P_1})}+\frac{1}{(a_0+\hat{P_2})})]
                 [\frac{(a_0+b)(a_0-b)}{[(a_0+\hat{Q_1})(a_0+\hat{Q_2})(a_0+\hat{P_1})
              (a_0+\hat{P_2})]^{\frac{1}{2}}}]$

$ C  =
\frac{1}{2}[\frac{2a_0}{a_0^2-b^2}-\frac{1}{2}(\frac{1}{(a_0+\hat{Q_1})}+\frac{1}{
              (a_0+\hat{Q_2})}+\frac{1}{(a_0+\hat{P_1})}+\frac{1}{(a_0+\hat{P_2})})]^2
[\frac{(a_0+b)(a_0-b)}{[(a_0+\hat{Q_1})(a_0+\hat{Q_2})(a_0+\hat{P_1})
              (a_0+\hat{P_2})]^{\frac{1}{2}}}]$

$  D  =
\frac{2(a_0+b)(a_0-b)}{[(a_0+\hat{Q_1})(a_0+\hat{Q_2})(a_0+\hat{P_1})
              (a_0+\hat{P_2})]^{\frac{1}{2}}}
                  [(\frac{1}{(a_0+\hat{Q_1})}+\frac{1}{
              (a_0+\hat{Q_2})}+\frac{1}{(a_0+\hat{P_1})}+\frac{1}{(a_0+\hat{P_2})})( \frac{1}{4}
              [\frac{1}{(a_0+\hat{Q_1})}+\frac{1}{
              (a_0+\hat{Q_2})}+\frac{1}{(a_0+\hat{P_1})}+\frac{1}{(a_0+\hat{P_2})}]^2 +
              \frac{-1}{2(a_0+\hat{Q_1})^2}+\frac{-1}{
              2(a_0+\hat{Q_2})^2}+\frac{-1}{2(a_0+\hat{P_1})^2}+\frac{-1}{2(a_0+\hat{P_2})^2})
              + \frac{1}{(a_0+\hat{Q_1})^3}+\frac{1}{
              (a_0+\hat{Q_2})^3}+\frac{1}{(a_0+\hat{P_1})^3}+\frac{1}{(a_0+\hat{P_2})^3}
              + ( \frac{-1}{2(a_0+\hat{Q_1})^2}+\frac{-1}{
              2(a_0+\hat{Q_2})^2}+\frac{-1}{2(a_0+\hat{P_1})^2}+\frac{-1}{2(a_0+\hat{P_2})^2})
              (\frac{1}{(a_0+\hat{Q_1})}+\frac{1}{
              (a_0+\hat{Q_2})}+\frac{1}{(a_0+\hat{P_1})}+\frac{1}{(a_0+\hat{P_2})}) ] $

$ E =
\frac{[\frac{2a_0}{a_0^2-b^2}-\frac{1}{2}(\frac{1}{(a_0+\hat{Q_1})}+\frac{1}{
              (a_0+\hat{Q_2})}+\frac{1}{(a_0+\hat{P_1})}+\frac{1}{(a_0+\hat{P_2})})]
                 [\frac{(a_0+b)(a_0-b)}{[(a_0+\hat{Q_1})(a_0+\hat{Q_2})(a_0+\hat{P_1})
              (a_0+\hat{P_2})]^{\frac{1}{2}}}]}{ (\frac{1}{2(a_0+\hat{Q_1})}+\frac{1}{
              2(a_0+\hat{Q_2})}+\frac{1}{2(a_0+\hat{P_1})}+\frac{1}{2(a_0+\hat{P_2})})}
              [(\frac{-1}{2(a_0+\hat{Q_1})^2}+\frac{-1}{
              2(a_0+\hat{Q_2})^2}+\frac{-1}{2(a_0+\hat{P_1})^2}+\frac{-1}{2(a_0+\hat{P_2})^2})
              +   \frac{1}{4}
              (\frac{1}{(a_0+\hat{Q_1})}+\frac{1}{
              (a_0+\hat{Q_2})}+\frac{1}{(a_0+\hat{P_1})}+\frac{1}{(a_0+\hat{P_2})})^2] $

$ G = 3 [\frac{1}{2(a_0+\hat{Q_1})}+\frac{1}{
              2(a_0+\hat{Q_2})}+\frac{1}{2(a_0+\hat{P_1})}+\frac{1}{2(a_0+\hat{P_2})}
              ]^2[\frac{(a_0+b)(a_0-b)}{[(a_0+\hat{Q_1})(a_0+\hat{Q_2})(a_0+\hat{P_1})
              (a_0+\hat{P_2})]^{\frac{1}{2}}}] $

$ H =
[\frac{(a_0+b)(a_0-b)}{[(a_0+\hat{Q_1})(a_0+\hat{Q_2})(a_0+\hat{P_1})
              (a_0+\hat{P_2})]^{\frac{1}{2}}}][(\frac{2a_0}{a_0^2-b^2}-\frac{1}{2}(\frac{1}{(a_0+\hat{Q_1})}+\frac{1}{
              (a_0+\hat{Q_2})}+\frac{1}{(a_0+\hat{P_1})}+\frac{1}{(a_0+\hat{P_2})}))^2
              - \frac{1}{(a_0+b)^2} - \frac{1}{(a_0-b)^2}+ \frac{1}{2}(\frac{1}{(a_0+\hat{Q_1})^2}+\frac{1}{
              (a_0+\hat{Q_2})^2}+\frac{1}{(a_0+\hat{P_1})^2}+\frac{1}{(a_0+\hat{P_2})^2})]     $

$ M =   2[(\frac{-1}{2(a_0+\hat{Q_1})^2}+\frac{-1}{
              2(a_0+\hat{Q_2})^2}+\frac{-1}{2(a_0+\hat{P_1})^2}+\frac{-1}{2(a_0+\hat{P_2})^2})
              + \frac{1}{4}(\frac{1}{(a_0+\hat{Q_1})}+\frac{1}{
              (a_0+\hat{Q_2})}+\frac{1}{(a_0+\hat{P_1})}+\frac{1}{(a_0+\hat{P_2})})^2]
              \sqrt{[\frac{(a_0+b)(a_0-b)}{[(a_0+\hat{Q_1})(a_0+\hat{Q_2})(a_0+\hat{P_1})
              (a_0+\hat{P_2})]^{\frac{1}{2}}}]}  [\frac{1}{2(a_0+\hat{Q_1})}+\frac{1}{
              2(a_0+\hat{Q_2})}+\frac{1}{2(a_0+\hat{P_1})}+\frac{1}{2(a_0+\hat{P_2})}
              ]  $

$ N =
\sqrt{[\frac{(a_0+b)(a_0-b)}{[(a_0+\hat{Q_1})(a_0+\hat{Q_2})(a_0+\hat{P_1})
              (a_0+\hat{P_2})]^{\frac{1}{2}}}]}[\frac{2a_0}{a_0^2-b^2}-\frac{1}{2}(\frac{1}{(a_0+\hat{Q_1})}+\frac{1}{
              (a_0+\hat{Q_2})}+\frac{1}{(a_0+\hat{P_1})}+\frac{1}{(a_0+\hat{P_2})})][\frac{1}{2(a_0+\hat{Q_1})}+\frac{1}{
              2(a_0+\hat{Q_2})}+\frac{1}{2(a_0+\hat{P_1})}+\frac{1}{2(a_0+\hat{P_2})}
              ]^2 $

$ L = [\frac{1}{2(a_0+\hat{Q_1})}+\frac{1}{
              2(a_0+\hat{Q_2})}+\frac{1}{2(a_0+\hat{P_1})}+\frac{1}{2(a_0+\hat{P_2})}
              ]^3 \sqrt{[\frac{(a_0+b)(a_0-b)}{[(a_0+\hat{Q_1})(a_0+\hat{Q_2})(a_0+\hat{P_1})
              (a_0+\hat{P_2})]^{\frac{1}{2}}}]}   $



\begin{thebibliography}{99}
\bibitem{kg6} M Visser, Nucl.Phys.B 328, 203 (1989)
\bibitem{kg10} W Israel, Nuovo Cimento 44B , 1 (1966) ; erratum -
ibid. 48B, 463 (1967)
        \bibitem{kg10} E Poisson and M Visser,
Phys.Rev.D 52, 7318 (1995) [arXiv: gr-qc / 9506083]

\bibitem{kg10} M Ishak and K Lake, Phys.Rev.D 65, 044011 (2002)
\bibitem{kg10} E Eiroa and G Romero,
Gen.Rel.Grav. 36, 651 (2004)[arXiv: gr-qc / 0303093]
\bibitem{kg10} F Lobo and P Crawford, Class.Quan.Grav. 21,
391 (2004)

\bibitem{kg10} E Eiroa and C Simeone,  Phys.Rev.D 70, 044008
(2004)

\bibitem{kg10} E Eiroa and C Simeone, Phys.Rev.D 71, 127501 (2005) [arXiv: gr-qc
/ 0502073]


\bibitem{kg10}  M Thibeault , C Simeone and E Eiroa,  gr-qc /
0512029

\bibitem{kg10}  F Rahaman, M Kalam and S Chakraborty, gr-qc/ 0607061
\bibitem{kg10} D Jatkar, S Mukherji and S Panda, Nucl. Phys. B 484,
223 ( 1997)
\bibitem{kg10} M Cvetic and D Youm,   Nucl. Phys. B 472,
249 ( 1996)
\bibitem{kg10} P Mitra,
Phys.Rev.D 57, 7369(1998)
\bibitem{kg10} A Sen,  Nucl.Phys.B 440, 421 (1995)
\bibitem{kg10} M Cvetic and D Youm,    Phys.Rev.D 53,
R584 ( 1996)
\end{thebibliography}
\end{document}